\documentclass[journal]{IEEEtran}
\usepackage{cite}
\usepackage{amsmath,amssymb,amsfonts}
\usepackage{graphicx}
\usepackage{subcaption}
\usepackage{textcomp}
\usepackage{url}
\def\BibTeX{{\rm B\kern-.05em{\sc i\kern-.025em b}\kern-.08em
    T\kern-.1667em\lower.7ex\hbox{E}\kern-.125emX}}

\begin{document}

\title{Classical Regularization in Variational Quantum Eigensolvers}

\author{%
  Yury~Chernyak$^{1*}$,
  Ijaz~Ahamed~Mohammad$^{1}$,
  and Martin~Plesch$^{1,2}$\\[0.5em]
  \small\textit{Manuscript received xxxx 00, 0000; revised xxxx 00, 0000.}\\[0.5em]
  \small$^{1}$Institute of Physics, Slovak Academy of Sciences, D\'ubravsk\'a cesta 5807/9, 84511 Bratislava, Slovakia.\\[0.5em]
  \small$^{2}$Matej Bel University, N\'arodn\'a ulica 12, 97401 Bansk\'a Bystrica, Slovakia.\\[0.5em]
  \small*Corresponding author: theofficialyury@gmail.com.
}

\markboth{Yury Chernyak et al.: Classical Regularization in Variational Quantum Eigensolvers}
{Yury Chernyak et al.: Classical Regularization in Variational Quantum Eigensolvers}

\maketitle

\begin{abstract}
While quantum computers are a very promising tool for the far future, in their current state of the art they remain limited both in size and quality. This has given rise to hybrid quantum--classical algorithms, where the quantum device performs only a small but vital part of the overall computation. Among these, variational quantum algorithms (VQAs), which combine a classical optimization procedure with quantum evaluation of a cost function, have emerged as particularly promising. However, barren plateaus and ill-conditioned optimization landscapes remain among the primary obstacles faced by VQAs, often leading to unstable convergence and high sensitivity to initialization. Motivated by this challenge, we investigate whether a purely classical remedy---standard L2 squared-norm regularization---can systematically stabilize hybrid quantum--classical optimization. Specifically, we augment the Variational Quantum Eigensolver (VQE) objective with a quadratic penalty $R(\boldsymbol{\theta}) = \lambda\|\boldsymbol{\theta}\|^2$, without modifying the quantum circuit or measurement process. Across all tested Hamiltonians---H$_2$, LiH, and the Random Field Ising Model (RFIM)---we observe improved performance over a broad window of the regularization strength $\lambda$. Our large-scale numerical results demonstrate that classical regularization provides a robust, system-independent mechanism for mitigating VQE instability, enhancing the reliability and reproducibility of variational quantum optimization without altering the underlying quantum circuit.
\end{abstract}


\section{Introduction}
\label{sec:introduction}
\IEEEPARstart{T}{he} Variational Quantum Eigensolver (VQE) has emerged as a central algorithm for estimating molecular and condensed-matter ground states on near-term quantum hardware. By coupling a parameterized quantum circuit with a classical optimizer, VQE minimizes an expectation value representing the system's energy. Yet despite this conceptual simplicity, practical implementations often display unstable convergence and high variability across random initializations—a reflection of the nonconvex and ill-conditioned nature of the hybrid optimization landscape. Among the most fundamental of these challenges is the occurrence of \emph{barren plateaus}~\cite{McClean2018BP,Grant2019AdiabaticBP,Cerezo2021CostFunctionDependentBP}, where gradients vanish exponentially with system size, rendering optimization effectively intractable for deep or highly entangled ansätze. Additional instabilities arise from optimizer sensitivity~\cite{Kandala2017HardwareEfficientVQE}, expressibility--trainability trade-offs~\cite{Sim2019ExpressibilityTrainability}, and finite-sampling noise~\cite{Arrasmith2021NoiseInducedBP}. Together, these challenges motivate strategies that can improve the conditioning of the classical optimization landscape without modifying the underlying quantum circuit.

Motivated by these instabilities, we explore a well-established technique from classical machine learning: quadratic $L_2^2$ regularization~\cite{Goodfellow2016DL,Ng2004FeatureSelection}. We incorporate a simple penalty
\[
R(\boldsymbol{\theta})=\|\boldsymbol{\theta}\|_2^2,
\]
scaled by a tunable coefficient~$\lambda$, directly into the classical objective. This \emph{classical regularization} leaves the quantum circuit unchanged while biasing the optimizer toward smoother and better-conditioned regions of parameter space.

We assess this approach across three representative systems—H$_2$, LiH, and the Random Field Ising Model (RFIM)—using a two-stage optimization protocol and thousands of random initializations. Our results reveal a universal non-monotonic dependence on~$\lambda$, with a broad stabilization window where convergence improves, parameter norms shrink, and success probabilities rise sharply. These findings demonstrate that a simple classical $L_2^2$ penalty can substantially enhance the reliability of VQE optimization without altering the quantum ansatz, offering a practical and broadly applicable improvement to hybrid quantum--classical algorithms.

\section{Conceptual and Algorithmic Background}

\subsection{Geometric Motivation for Regularization}

Variational quantum eigensolvers (VQEs) often employ expressive ansätze with many tunable parameters. While such flexibility is desirable, it also introduces an abundance of redundant degrees of freedom: different parameter configurations can generate quantum states that are physically identical or nearly so. When viewed as a function of the parameters~$\theta$, the classical energy landscape inherits this redundancy in the form of shallow valleys, sharp ridges, and rapidly oscillatory curvature. As a result, even well-behaved physical Hamiltonians can give rise to unstable or erratic optimization trajectories.

Regularization enters this picture not through its statistical interpretation, but through its geometric effect. In classical nonlinear optimization, $L_2^2$ penalties improve conditioning by shrinking parameter directions that contribute little to the underlying model yet produce large or rapidly varying curvature in the objective. Importantly, this stabilizing mechanism requires neither a dataset nor a notion of generalization. Instead, it reshapes the geometry of the objective function, biasing the search toward smoother and better-conditioned regions of parameter space.

An analogous opportunity arises in VQE. Because many~$\theta$-directions influence the quantum state only weakly, yet strongly affect the curvature of the classical energy surface, these directions behave as ``oscillatory modes'' that complicate gradient-based search. Adding an $L_2^2$ penalty suppresses excursions into such ill-conditioned directions without altering the physical location of the ground-state minimum. This geometric perspective motivates our investigation: can classical regularization improve the stability and reliability of variational quantum optimization purely through landscape conditioning?

\subsection{Regularization in ML}

Although $L_2^2$ penalties originate in machine learning, their classical purpose is fundamentally statistical. In supervised learning, a model~$f_\theta(x)$ is fitted to a dataset~$\{(x_i, y_i)\}_{i=1}^N$ by minimizing the empirical loss
\[
L(\theta) = \sum_{i=1}^N \ell\!\left(f_\theta(x_i), y_i\right).
\]
When the number of parameters~$P$ greatly exceeds the number of datapoints~$N$, the system becomes underdetermined: many values of~$\theta$ interpolate the data exactly, i.e.,
\[
f_\theta(x_i) = y_i \quad \text{for all } i.
\]
Among these infinitely many exact-fit solutions are highly oscillatory or overly complex functions that perfectly match the training data but perform poorly on unseen inputs—a failure of generalization~\cite{belkin2019}.

$L_2^2$ regularization addresses this issue by modifying the optimization problem to
\[
L_{\mathrm{reg}}(\theta) = L(\theta) + \lambda \|\theta\|^2,
\]
which selects the minimum-norm solution among all exact fits. In linear models, this minimum-norm corresponds to the smoothest function consistent with the training data, and in nonlinear models the same mechanism suppresses unnecessary degrees of freedom, steering the solution toward smoother functions while reducing the high variance associated with overparameterized models~\cite{hastie2009,bishop2006}. This role of $L_2^2$ penalties is therefore deeply statistical: it balances fit to the data with model complexity to promote good generalization~\cite{belkin2019}.

\subsection{Regularization in VQE}

VQE operates in a fundamentally different regime. There is no dataset, no distribution of unseen inputs, and no bias–variance tradeoff. The objective is a single scalar quantity—the energy expectation value
\[
\langle \psi(\theta) | H | \psi(\theta) \rangle,
\]
constructed from a single parametrized quantum state. Regularization in VQE thus plays no role in controlling model complexity or preventing overfitting.

Instead, its effect is purely geometric. Expressive variational ansätze contain redundant parameter directions: many distinct~$\theta$ produce quantum states that are physically identical or nearly so~\cite{sim2019}. These directions generate large or rapidly oscillating curvature in the classical energy landscape~\cite{mcclean2018, cerezo2021} while contributing negligibly to the energy itself. Adding an $L_2^2$ penalty suppresses excursions into such ill-conditioned directions, improving the conditioning and stability of the optimization without restricting the expressive power of the ansatz.

This geometric role of $L_2^2$ regularization is entirely distinct from its statistical role in machine learning. Yet the same mathematical mechanism—shrinking parameter values in directions that destabilize the objective—motivates its use in VQE.

\section{Regularized VQE Framework}

\subsection{Regularized Objective}
The Variational Quantum Eigensolver (VQE) seeks to minimize the expectation value
\(
E(\boldsymbol{\theta}) = \langle \psi(\boldsymbol{\theta}) | \hat{H} | \psi(\boldsymbol{\theta}) \rangle,
\)
where $\boldsymbol{\theta}$ parameterizes the quantum circuit and $\hat{H}$ is the target Hamiltonian.
In practice, however, this optimization landscape is often rugged and poorly conditioned, hindering convergence.
To address this challenge, we introduce a classical quadratic penalty on the parameters, yielding the regularized objective
\begin{equation}
  \tilde{E}(\boldsymbol{\theta}) = E(\boldsymbol{\theta}) + \lambda R(\boldsymbol{\theta}),
  \qquad R(\boldsymbol{\theta}) = \|\boldsymbol{\theta}\|_2^2,
  \label{eq:objective}
\end{equation}
where $\lambda \ge 0$ controls the regularization strength.

Conceptually, this penalty suppresses parameter directions that contribute little to lowering the energy yet inflate the overall norm,
thereby smoothing sharp valleys and flattening ill-conditioned ridges in the cost surface.
By encouraging trajectories to remain within a compact and physically meaningful region of parameter space,
this conditioning reduces sensitivity to random initialization and prevents divergent oscillations.
Crucially, unlike circuit-based techniques that modify the quantum layer itself, this purely classical regularizer reshapes only the optimizer's search geometry—making it compatible with any variational ansatz or backend.

\subsection{Systems and Hamiltonians}
To implement this regularization strategy, all simulations were performed using the open-source framework on GitHub~\cite{Chernyak2025Pipeline}.
This modular environment allows plug-and-play replacement of Hamiltonians, classical optimizers, regularization schedules, and ansatz definitions,
enabling systematic comparisons under unified evaluation and logging tools.

To systematically explore the effects of regularization across diverse optimization landscapes, we selected three representative systems, each chosen to probe distinct energy-landscape characteristics:
\begin{itemize}
  \item \textbf{H$_2$ molecule}: A four-qubit system with a 40-parameter hardware-efficient \textit{TwoLocal} ansatz (four layers), serving as a well-characterized baseline for molecular VQE. The number of parameters arise from two rotation blocks per qubit per layer (Ry and Rz) across four repetitions, plus an initial rotation layer.
  \item \textbf{LiH molecule}: An eight-qubit system with 80 parameters using the same ansatz topology, chosen to evaluate how regularization scales with circuit dimensionality.
  \item \textbf{Random Field Ising Model (RFIM)}: A 12-qubit system with random on-site fields and a single-layer Ry ansatz (12 parameters), providing a non-chemical benchmark characterized by rugged, strongly nonconvex energy landscapes~\cite{RFIM}.
\end{itemize}

\begin{figure}[htbp]
  \centering
  \includegraphics[width=0.98\linewidth]{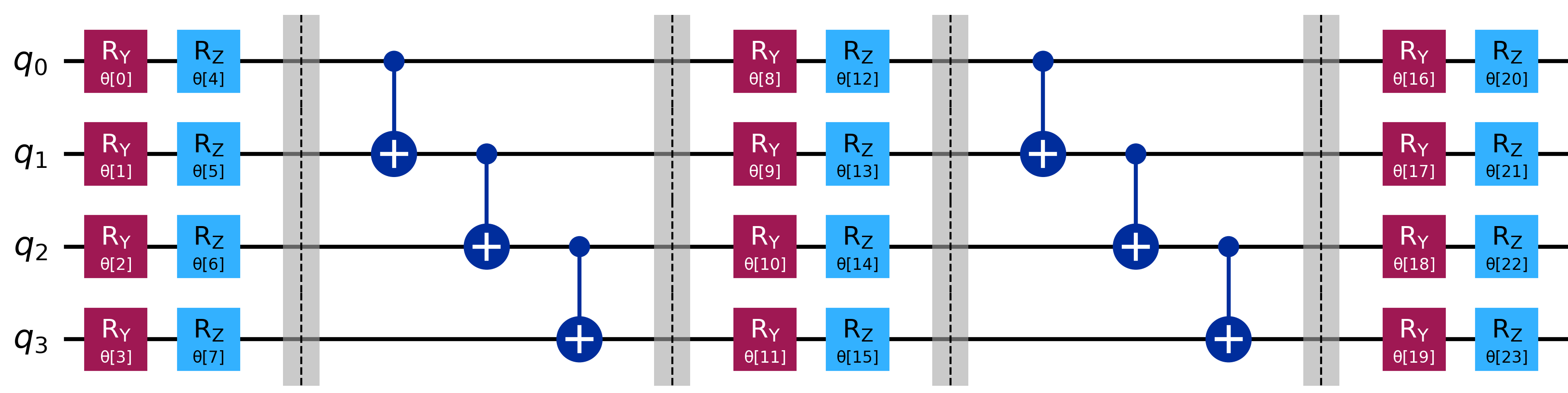}
  \caption{Example of the 4-qubit hardware-efficient \textit{TwoLocal} ansatz used for H$_2$ (Ry/Rz blocks). The experiment used an initial rotation layer followed by 4 layers, meanwhile in this figure only 2 layers are depicted for optimal visual clarity.}
  \label{fig:h2_ansatz}
\end{figure}

For consistency across all experiments, Hamiltonians were encoded as weighted Pauli-sum operators via Qiskit's SparsePauliOp class.
To isolate purely algorithmic behavior from hardware noise, each system was simulated under noiseless conditions using Qiskit's \textit{StatevectorEstimator} backend~\cite{Qiskit2024}, which computes exact expectation values via matrix operations rather than sampling.

\subsection{Optimizers}
To disentangle the stabilizing effects of regularization from final convergence quality, all experiments employed a two-stage optimization structure.
In the first phase (Stage~A), a regularized exploration promotes stable search; in the second phase (Stage~B), an unregularized refinement polishes the solution.
The total function-evaluation budget was shared between these two stages.

For the H$_2$ and RFIM Hamiltonians, we used the Conjugate Gradient (CG) method implemented in scipy.optimize class, configured with a tolerance of $10^{-2}$ and iteration limits of 15 for Stage~A and 10 for Stage~B.
This yielded an approximate evaluation budget of 10,000 function calls per configuration (i.e., per combination of Hamiltonian, optimizer, and $\lambda$ value).
For the larger LiH system, we employed the limited-memory BFGS optimizer (L-BFGS-B) with a higher iteration cap of 200 per stage to accommodate its increased dimensionality.
Across all systems, the cosine-decay parameter $T$ in the regularization schedule was matched to the Stage-A iteration count,
ensuring that~$\lambda$ decayed smoothly to zero before transitioning to the unregularized phase.
This design guarantees consistent temporal alignment between the scheduler and the optimization horizon.

\subsection{Two-Stage Regularization with Cosine Scheduler}
To isolate and quantify the stabilizing role of classical regularization, we devised a two-stage optimization scheme that separates exploration from refinement.
In the first phase (Stage~A), the regularization term in Eq.~\eqref{eq:objective} remains active and follows a smooth cosine decay:
\begin{equation}
  \lambda(t) = \frac{\lambda_0}{2}\!\left[1 + \cos\!\left(\frac{\pi t}{T_A}\right)\right],
  \qquad 0 \le t \le T_A,
\end{equation}
where $\lambda_0$ is the initial regularization strength and $T_A$ the number of iterations in Stage~A.
This cosine-decay schedule ensures a smooth transition between exploration and refinement, gradually reducing the penalty as the optimizer approaches a stable region.
After this period, $\lambda(t)$ smoothly approaches zero, completing the first stage of regularized optimization.
The second phase (Stage~B) then resumes from the best parameters of Stage~A with $\lambda=0$, performing an unregularized fine-tuning of the minimum found under the stabilizing prior.
This two-stage design allows us to measure the effect of regularization during exploration while ensuring that the final energy estimates remain unbiased by the penalty term.

\subsection{Finding the Optimal Range of $\lambda$}
To identify the regularization strength that maximizes robustness without sacrificing accuracy, we systematically scanned $\lambda_0$ values across multiple orders of magnitude, typically in the interval
\[
\begin{split}
\lambda_0 \in \{&0.00, 0.005, 0.020, 0.050, 0.075, 0.099, \\
&0.125, 0.150, 0.175, 0.200, 0.250\}.
\end{split}
\]
We explored each setting over thousands of random initializations—10,000 runs for H$_2$, 6,000 for LiH, and 10,000 for RFIM—recording success rates (fraction of runs reaching target accuracy), median final energies, and parameter-norm statistics at each $\lambda_0$ value.
This data-driven approach defines $\lambda_{\mathrm{opt}}$ as the contiguous range of $\lambda$ values for which the mean success rate remains at least 90\,\% of the global maximum upon reaching a defined target accuracy, thereby identifying the window where regularization provides consistent benefit across random initializations. The 90\% threshold is chosen as a robustness criterion rather than a sharp optimum, ensuring that $\lambda_{\mathrm{opt}}$ captures a stable performance plateau rather than isolated peak values; qualitatively similar windows are obtained for nearby thresholds.

\subsection{Evaluation Metrics}
To ensure comprehensive assessment of regularization effects, we tracked four key metrics during all experiments:
\begin{enumerate}
  \item Final Stage-B energies $E_\mathrm{final}$ to assess ground-state convergence.
  \item Per-run parameter norms $\|\boldsymbol{\theta}\|_2$ as indicators of damping.
  \item Success rates under chemical-accuracy thresholds
  $\Delta E \le \{1.5\times10^{-1},\,1.5\times10^{-2},\,\dotsc,\,1.5\times10^{-7}\}$\,Ha, computed at 95\,\% confidence using the Wilson interval~\cite{Wilson1927}.
  \item Iteration-wise trajectories of energy, objective, and parameter norm.
\end{enumerate}
These metrics collectively reveal both the final quality of solutions and the dynamics of convergence.

A run is counted as successful if the final VQE energy lies within $\delta$ of the exact ground-state energy.
We evaluate several chemical-accuracy thresholds (as defined in Section IV-F), but the cross-system comparisons in the main text use the threshold shown in each figure caption.
This definition allows us to quantify the fraction of random initializations that converge to a physically meaningful solution under varying regularization strengths.
Statistical summaries were produced using Numpy and Pandas, while visualization dashboards were generated with Plotly scripts (v5.x) in the analysis/create\_dashboard.py module~\cite{Chernyak2025Pipeline, Plotly2024}.

\subsection{Computational Setup}
All experiments were executed with 15 parallel processes (16 CPUs total), using Python~3.13, NumPy~2.3, SciPy~1.16, and Qiskit~2.1.
Each simulation employed the noiseless \textit{StatevectorEstimator} to isolate algorithmic effects from sampling noise.
Additional large-scale sweeps were distributed via SLURM on the Devana supercomputer for the complex LiH 80-parameter system.
\label{sec:methods}

\section{Results}

The large-scale experiments performed on H$_2$, LiH, and the Random Field Ising Model (RFIM) reveal a consistent and striking pattern. Across all systems, the addition of a classical $L_2^2$ regularizer stabilizes convergence and improves the reliability of optimization outcomes in a systematic manner. The success-rate curves as a function of $\lambda$ exhibit a clear nonmonotonic trend: performance improves as $\lambda$ increases from zero, reaches an optimal regime at moderate regularization strength, and degrades as the regularization becomes overly restrictive. This behavior identifies a distinct stabilization window in which search damping enhances convergence without compromising expressivity.

Across all three systems, this stabilization window is consistently observed, indicating that the effect is not system-specific. The detailed energy distributions, parameter-norm evolution, and optimization trajectories under different regularization strengths provide further insight into the mechanisms underlying this stabilization.

\subsection{H$_2$ Molecule}
\begin{figure*}[t]
  \centering

  % ---------- Row 1 ----------
  \begin{subfigure}[b]{0.47\textwidth}
    \centering
    \includegraphics[width=\textwidth]{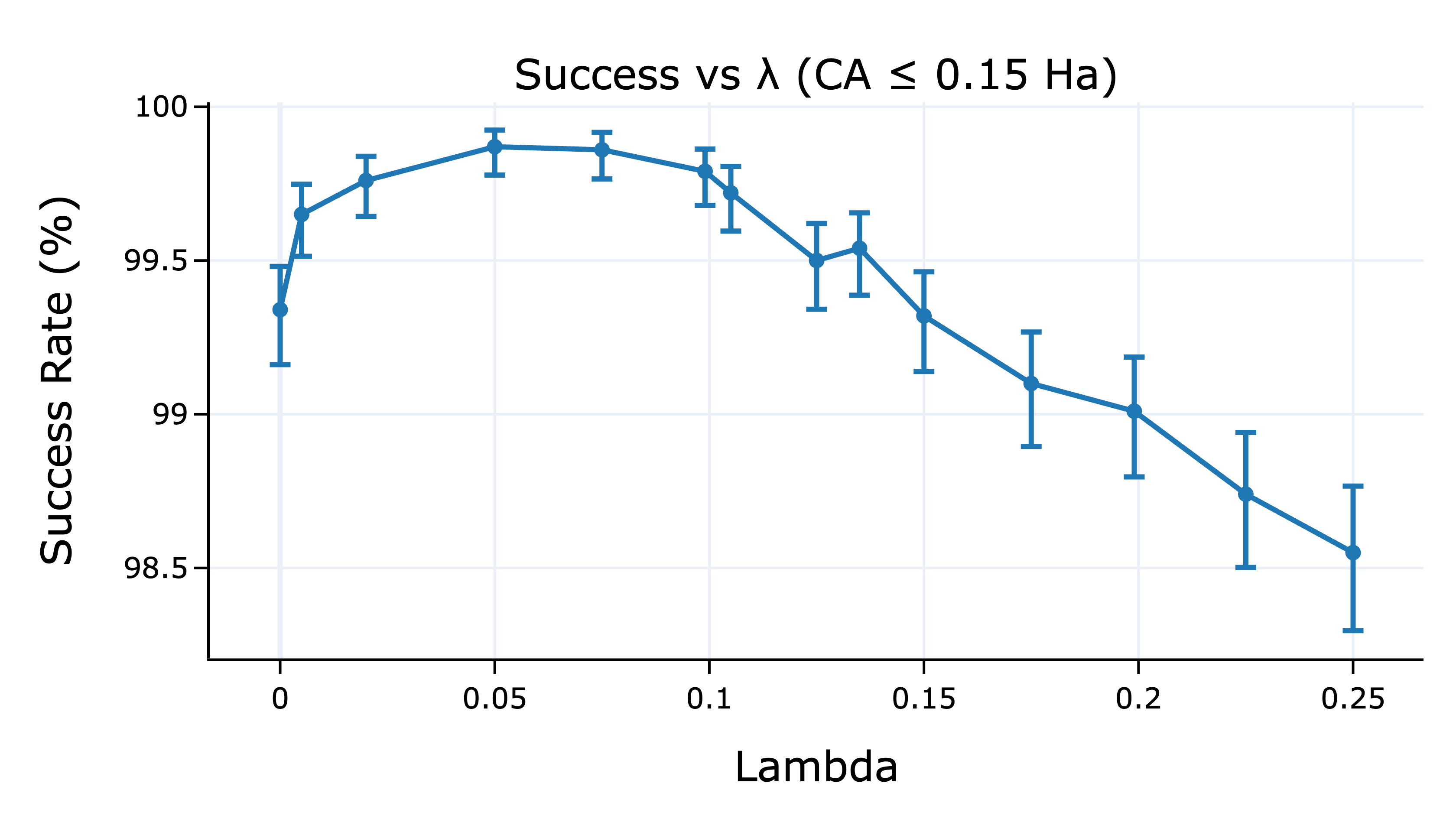}%
    \caption{}
    \label{fig:CA_0.15_h2}
  \end{subfigure}\hfill
  \begin{subfigure}[b]{0.47\textwidth}
    \centering
    \includegraphics[width=\textwidth]{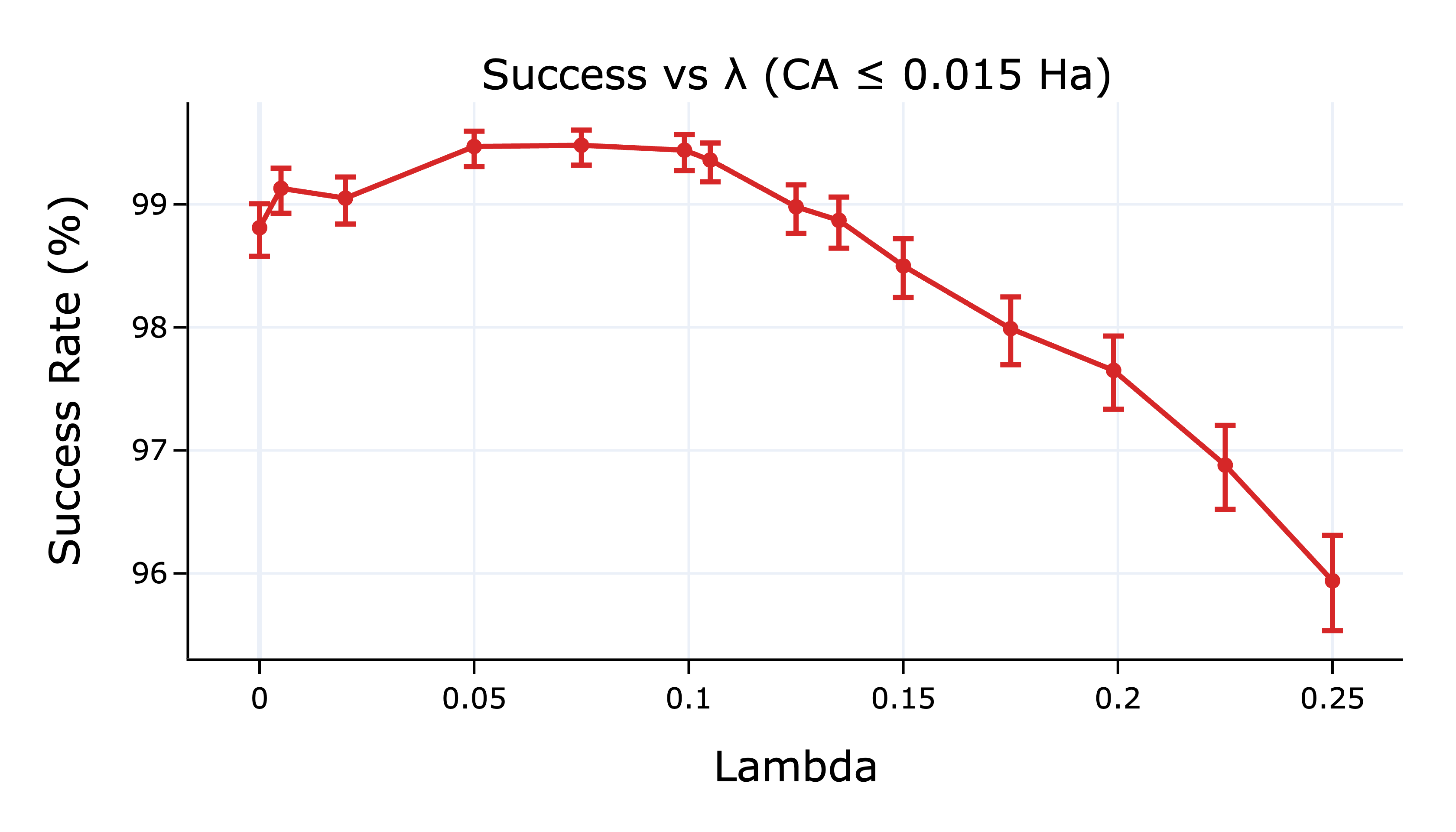}%
    \caption{}
    \label{fig:CA_0.015_h2}
  \end{subfigure}

  \vspace{0.4em}

  % ---------- Row 2 ----------
  \begin{subfigure}[b]{0.47\textwidth}
    \centering
    \includegraphics[width=\textwidth]{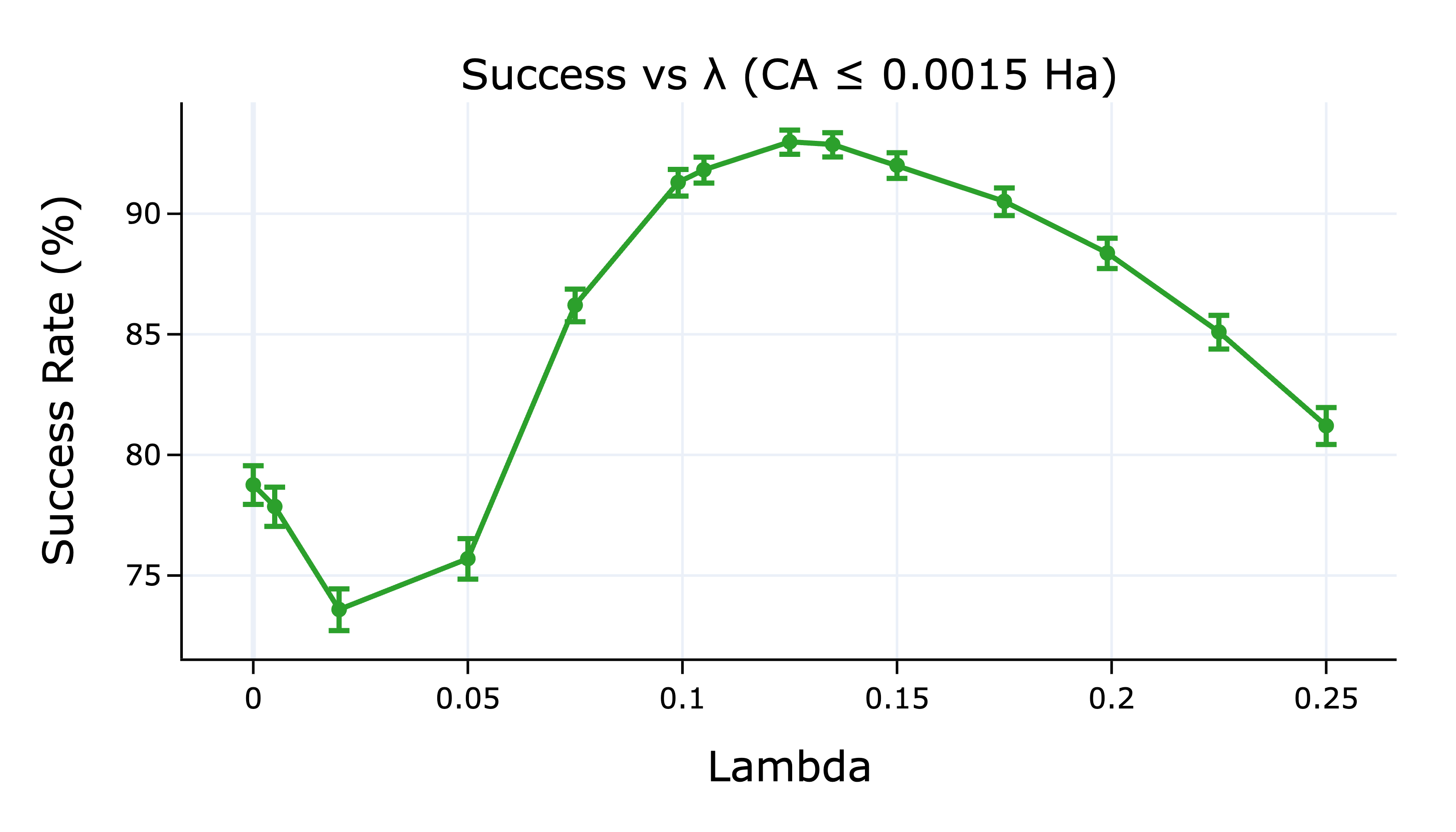}%
    \caption{}
    \label{fig:CA_0.0015_h2}
  \end{subfigure}\hfill
  \begin{subfigure}[b]{0.47\textwidth}
    \centering
    \includegraphics[width=\textwidth]{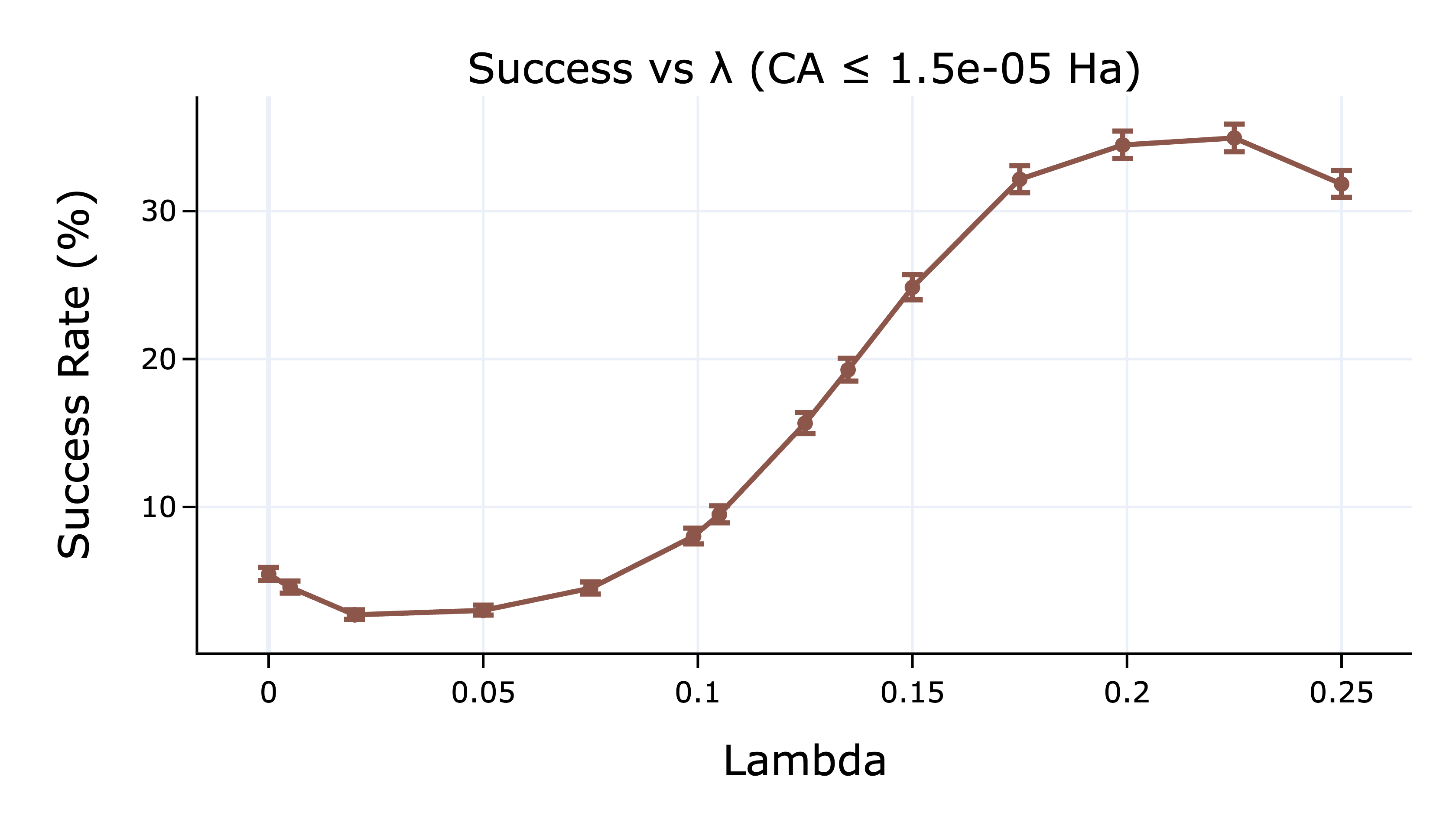}%
    \caption{}
    \label{fig:CA_1.5E-05_h2}
  \end{subfigure}

  \vspace{0.4em}

  % ---------- Row 3 ----------
  \begin{subfigure}[b]{0.47\textwidth}
    \centering
    \includegraphics[width=\textwidth]{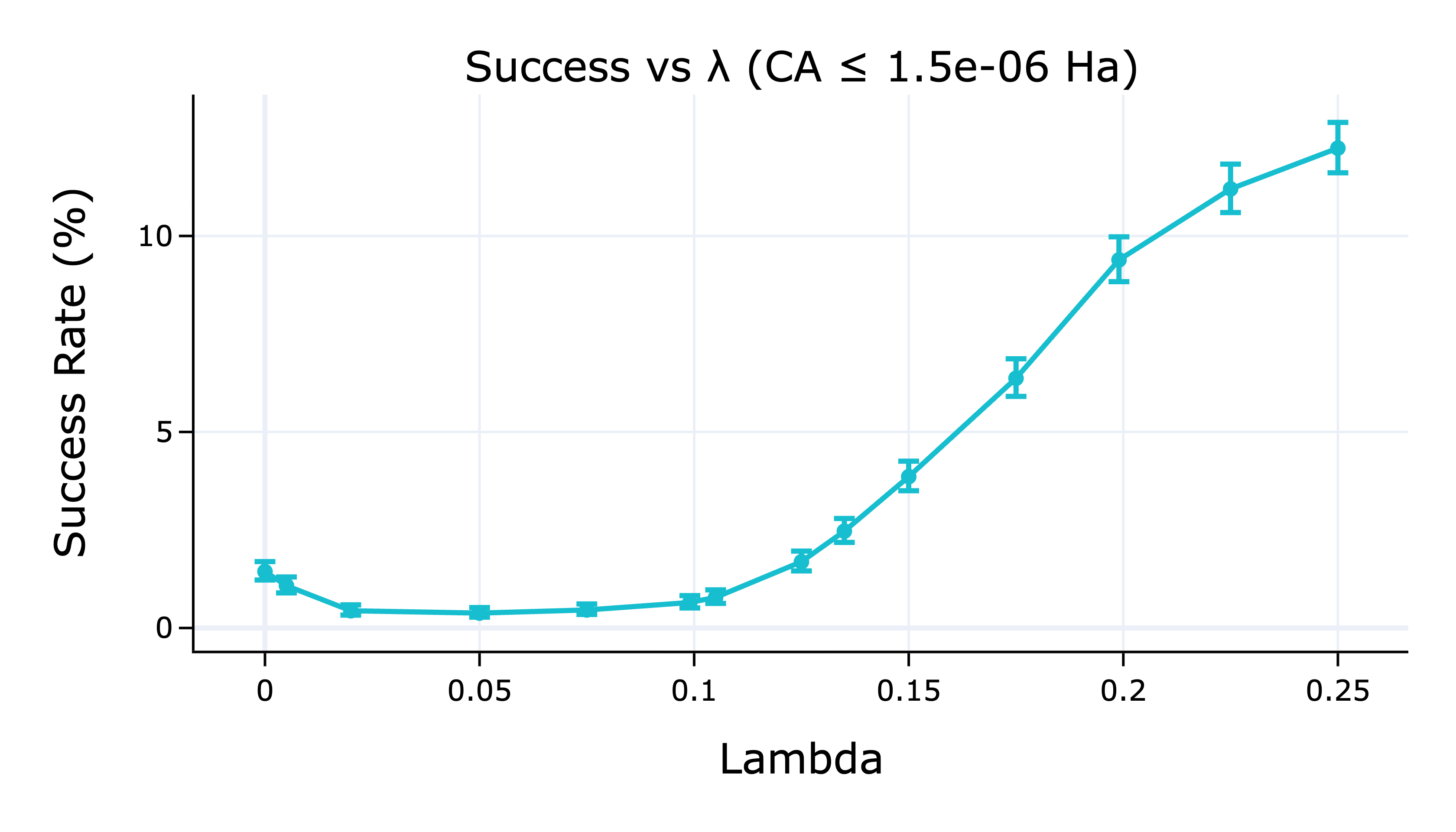}%
    \caption{}
    \label{fig:CA_1.5E-06_h2}
  \end{subfigure}\hfill
  \begin{subfigure}[b]{0.47\textwidth}
    \centering
    \includegraphics[width=\textwidth]{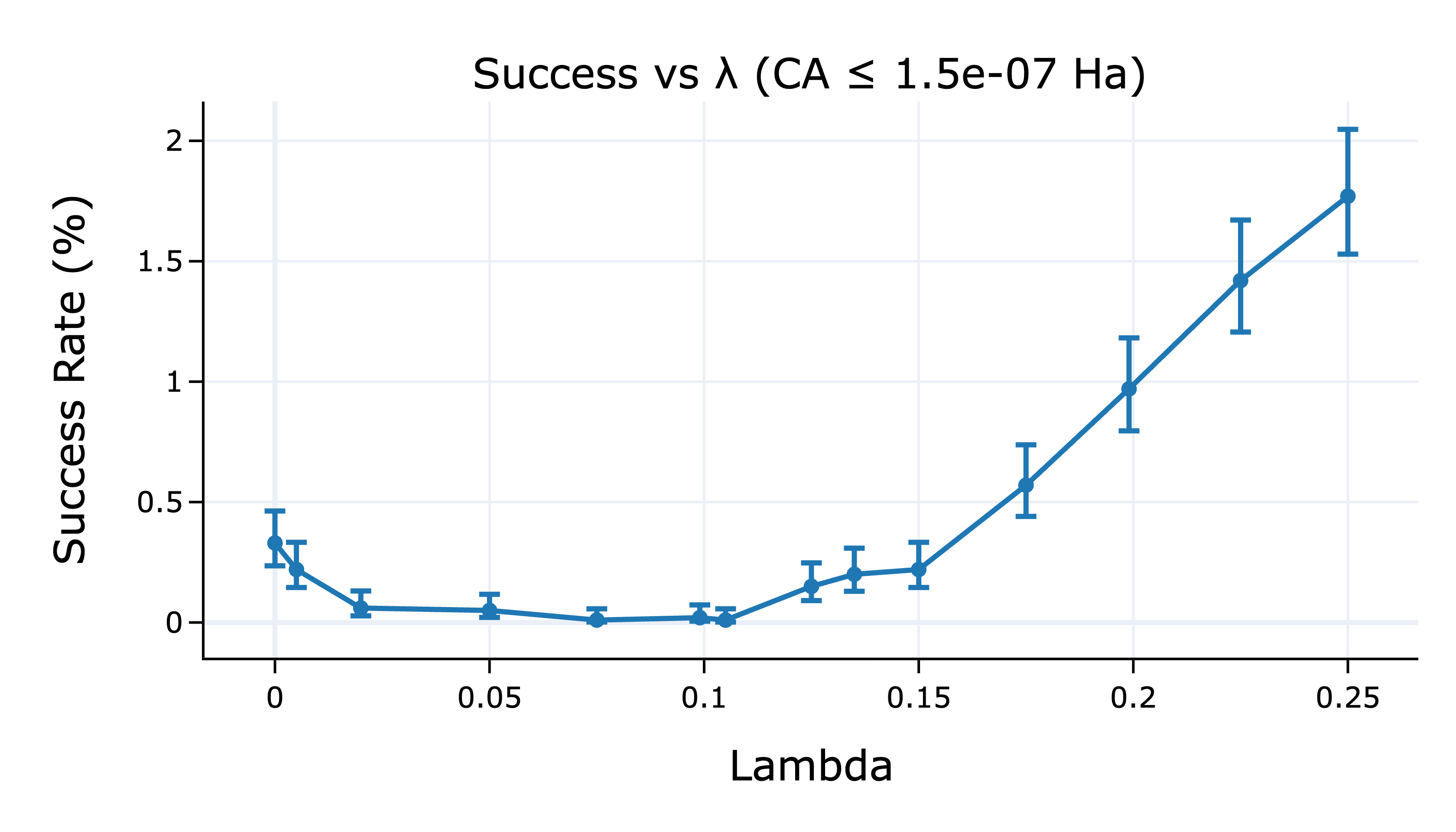}%
    \caption{}
    \label{fig:CA_1.5E-07_h2}
  \end{subfigure}

  \caption{H$_2$ molecule: success rate versus~$\lambda$ for different
  chemical-accuracy thresholds. Panels (a)–(f) show the fraction of runs
  reaching $\Delta E \le 1.5\times10^{-k}$ Ha for decreasing values of $k$.
  A broad moderate-regularization window yields consistently high success
  rates across thresholds.}
  \label{fig:h2-success-grid}
\end{figure*}

The hydrogen molecule serves as the baseline for evaluating the influence of regularization, offering a well-characterized test case with a 40-parameter hardware-efficient ansatz.
We explored each configuration with 10,000 random seeds per~$\lambda$ under noiseless statevector estimation, providing robust statistical power to detect even subtle effects.
As shown in Figure~\ref{fig:h2-success-grid}, the success rate increased sharply with small~$\lambda$ and reached its maximum in the interval $\lambda\approx0.10$--$0.18$.
Beyond $\lambda\gtrsim0.2$, performance declined gradually as the penalty began to dominate the cost landscape, demonstrating the characteristic nonmonotonic dependence.
It is interesting to note that, with stricter accuracy thresholds, the peak success rate shifts to a higher value of $\lambda$. This is visible in Fig.~2: the curve with chemical-accuracy threshold $\mathrm{CA} \leq 0.15\,\mathrm{Ha}$ peaks around $\lambda \approx 0.05$, whereas the curve with $\mathrm{CA} \leq 1.5\times10^{-7}\,\mathrm{Ha}$ peaks near $\lambda \approx 0.25$.

The benefits of moderate regularization manifest clearly in the energy statistics.
The interquartile range of final Stage-B energies contracted by approximately 35--40\,\%, while parameter-norm spreads decreased by a comparable amount, indicating substantial tightening of the energy distribution.
Crucially, the mean final energy remained statistically indistinguishable from the unregularized baseline, confirming that the regularizer improved conditioning without biasing the optimum.
Trajectory analysis further revealed a smoother decay of both energy and parameter norm during Stage~A, followed by rapid convergence during Stage~B—evidence that the penalty guides exploration toward well-conditioned regions before refinement begins.

We also observe a shallow dip in success probability at very small but nonzero values of $\lambda$, most clearly visible in Fig.~2(c)--(f). In this regime, the regularization term is sufficiently weak that it does not fully suppress unstable directions, yet strong enough to perturb early optimization trajectories. As a result, a small fraction of runs that would otherwise converge successfully in the unregularized case are temporarily diverted, leading to a modest reduction in success rate. Importantly, this dip disappears as $\lambda$ increases into the stabilization window, indicating that it reflects a transitional regime rather than a failure of regularization.

\subsection{LiH Molecule}
To test whether this stabilization pattern generalizes to higher-dimensional systems, we extended the analysis to the LiH molecule, which employs an 80-parameter ansatz across eight qubits.
Each $\lambda$ value was tested over approximately 6,000 random seeds, providing sufficient statistical power to resolve systematic trends across regularization strengths.
For LiH, optimization was performed using the L-BFGS algorithm, in contrast to the conjugate-gradient optimizer used for the H$_2$ and RFIM instances.
The overall shape of the success curve mirrored that of H$_2$, though the optimal window shifted to smaller $\lambda$ values---specifically $\lambda\approx0.005$--$0.025$.

The downward shift of the optimal $\lambda$ window relative to H$_2$ is consistent with the increased dimensionality of the LiH ansatz. With twice as many variational parameters, it is observed that stabilization occurs at smaller regularization strengths. This seems to indicate an inverse relationship between $\lambda_{\mathrm{opt}}$ and parameter count.

Regularization consistently reduced variance in both energies and parameter norms.
At very small~$\lambda$, several runs diverged or stagnated, whereas moderate~$\lambda$ yielded smooth and reproducible convergence.
Overly large~$\lambda$ values ($>0.05$) led to slower convergence and a slight upward bias in the final energies, indicating excessive damping.
These results confirm that the stabilizing window narrows and shifts downward with increasing system dimension.

\begin{figure}[t]
  \centering

  \begin{subfigure}[b]{\linewidth}
    \centering
    \includegraphics[width=\linewidth]{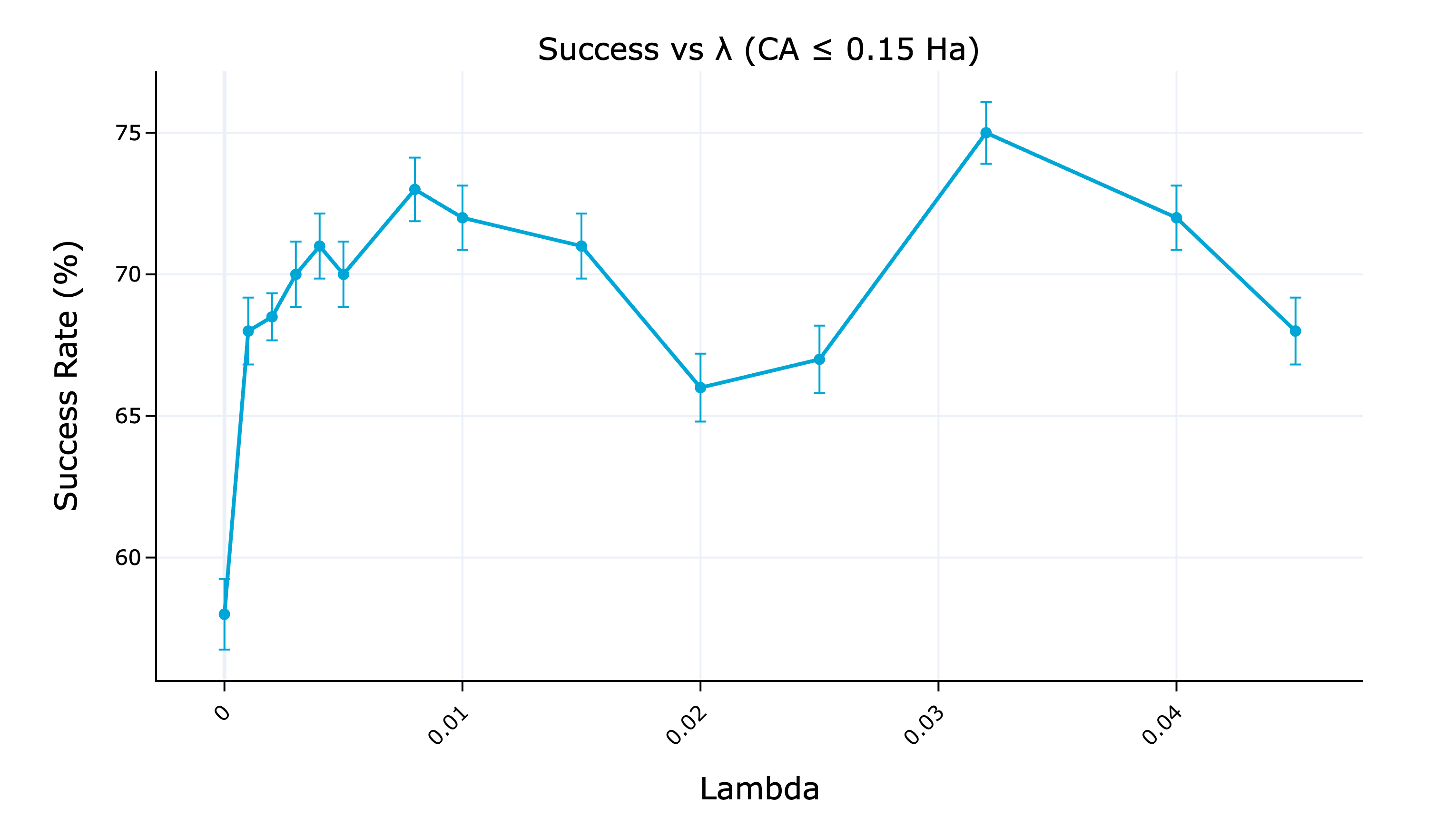}
    \caption{}
    \label{fig:lih-ca015}
  \end{subfigure}

  \vspace{0.6em}

  \begin{subfigure}[b]{\linewidth}
    \centering
    \includegraphics[width=\linewidth]{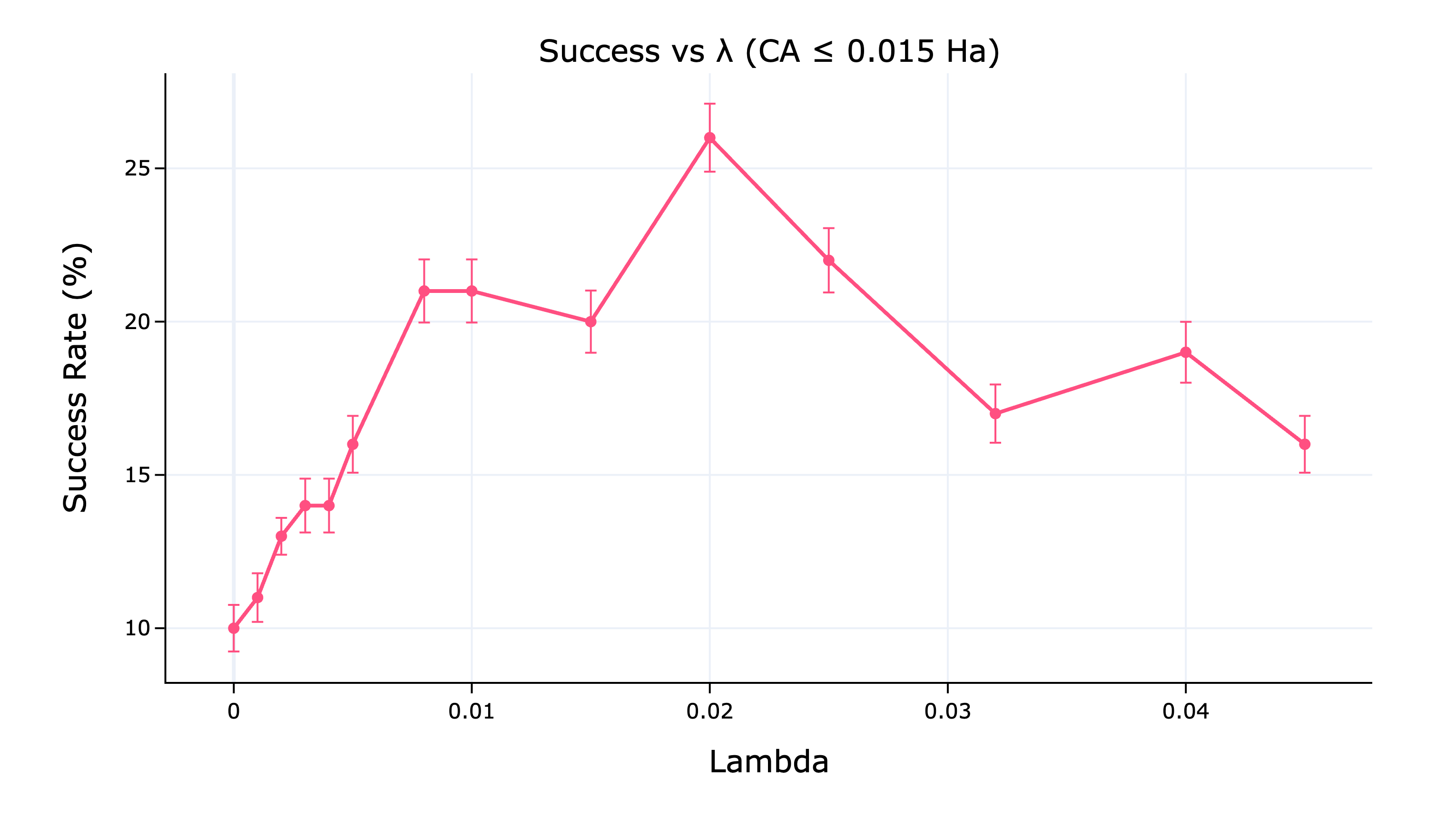}
    \caption{}
    \label{fig:lih-ca0015}
  \end{subfigure}

  \caption{LiH molecule: success rate versus~$\lambda$ for two chemical-accuracy thresholds. The stabilizing window shifts to smaller $\lambda$ values compared with H$_2$, reflecting increased dimensionality and stronger redundant directions in the ansatz.}
  \label{fig:lih_stats}
\end{figure}

\subsection{Random Field Ising Model (RFIM)}
To determine whether the observed stabilization depends on chemical structure or arises from more general optimization dynamics, we turned to the Random Field Ising Model---a 12-qubit Hamiltonian with no entanglement involved and characterized by strong nonconvexity and numerous local minima.
Each configuration was executed with 10,000 random seeds per~$\lambda$, ensuring robust statistics across this challenging landscape.
Despite its rugged energy surface, the same qualitative pattern emerged as in the previous cases, demonstrating that regularization stabilizes optimization independently of the underlying physical system.
Moderate regularization substantially improved convergence reliability and suppressed oscillatory updates during the CG iterations.
Optimal $\lambda$ values lay in the broad interval $0.05$--$0.12$, producing roughly double the success probability of the unregularized baseline.
At larger~$\lambda$, trajectories became over-damped and the achieved minima shifted upward in energy, again reflecting an over-constrained regime.

\begin{figure}[t]
  \centering

  \begin{subfigure}[b]{\linewidth}
    \centering
    \includegraphics[width=\linewidth]{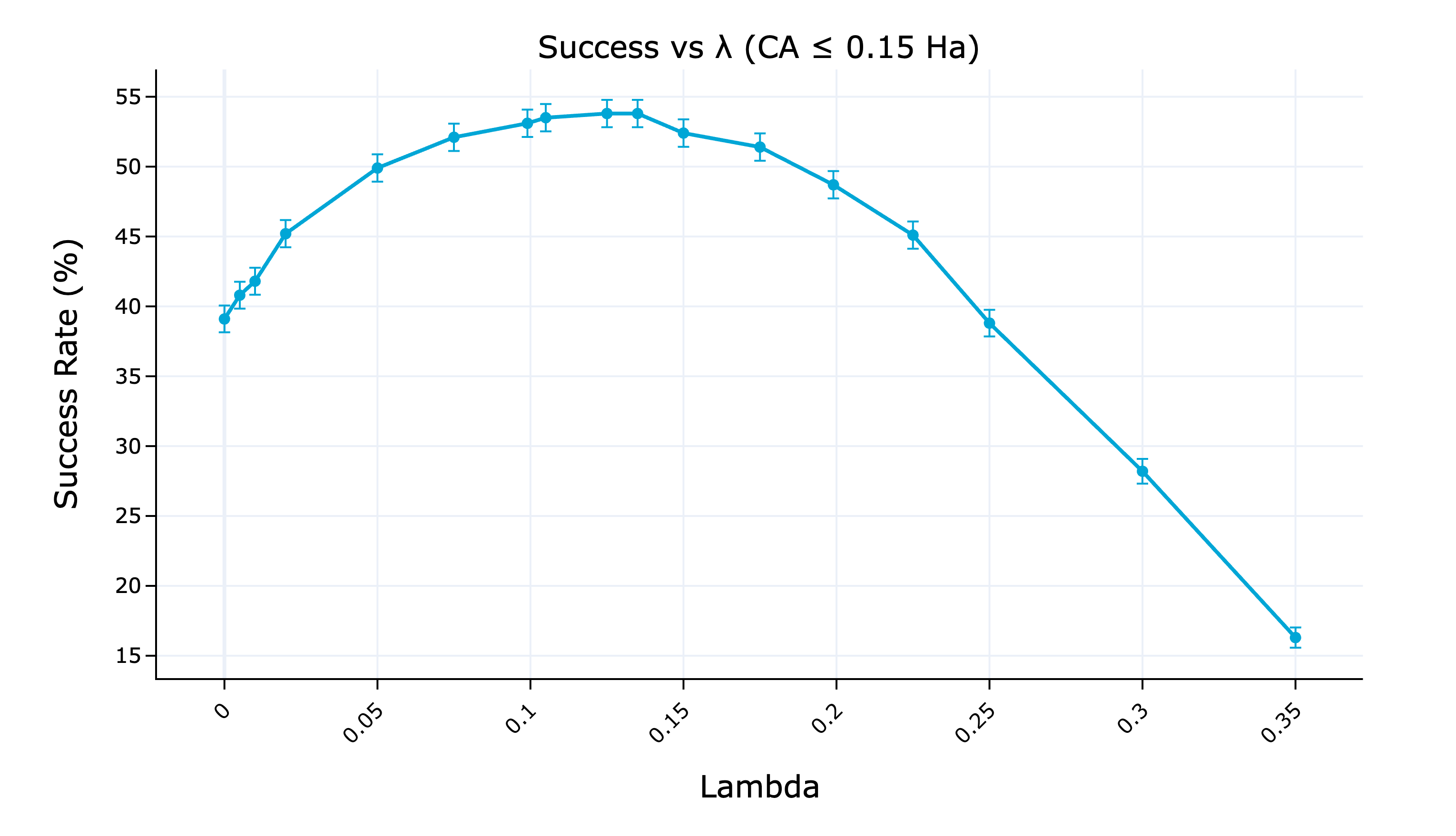}
    \caption{}
    \label{fig:rfim-ca015}
  \end{subfigure}

  \vspace{0.6em}

  \begin{subfigure}[b]{\linewidth}
    \centering
    \includegraphics[width=\linewidth]{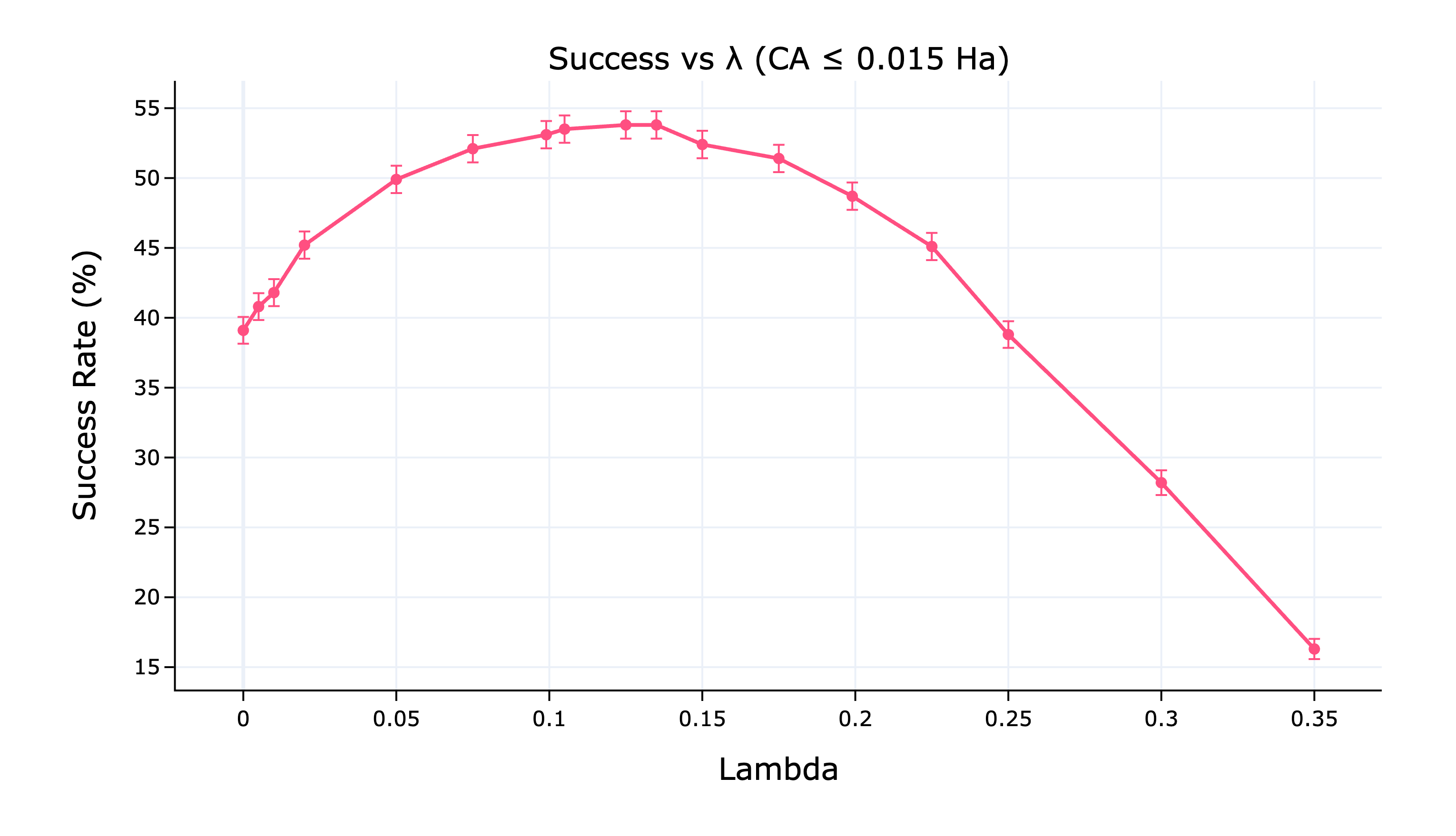}
    \caption{}
    \label{fig:rfim-ca0000015}
  \end{subfigure}

  \caption{RFIM benchmark: success rate versus~$\lambda$ for two chemical-accuracy thresholds. The stabilizing window lies in the broad interval $\lambda \approx 0.05$--$0.12$, and moderate regularization consistently improves convergence reliability across this strongly nonconvex landscape.}
  \label{fig:rfim_stats}
\end{figure}

\subsection{Cross-System Trends}
Having analyzed each system individually, we now synthesize these findings to identify universal stabilization trends.
Across all studied Hamiltonians, the qualitative impact of classical regularization followed the same three-phase pattern:
\begin{enumerate}
  \item Low $\lambda$ values correspond to unconstrained searches with large variance and high sensitivity to initialization.
  \item Moderate $\lambda$ values yield stabilized convergence, higher success probability, and tighter energy distributions.
  \item Excessively large $\lambda$ values over-constrain the optimization and reduce attainable accuracy.
\end{enumerate}

Together, Figures~\ref{fig:h2-success-grid}--\ref{fig:rfim_stats} illustrate that the stabilizing effect of $\lambda$ is consistent across all systems, with each exhibiting the same nonmonotonic trend and an identifiable stabilization window.

To provide a system-aware reference scale for selecting the regularization strength $\lambda$, we consider the overall magnitude of the Hamiltonian
\[
H = \sum_i c_i P_i,
\]
where $P_i$ denote Pauli operators, and normalize this scale by the number of variational parameters $P$. This yields the heuristic estimate
\[
\lambda_{\text{scale}} \sim \frac{\sum_i |c_i|}{P}.
\]
We use $\lambda_{\text{scale}}$ as an order-of-magnitude reference for selecting the regularization strength. Rather than predicting an optimal value, this scale indicates where regularization effects become relevant. In practice, evaluating $\lambda_{\text{scale}}$ together with values a factor of a few larger and smaller (e.g., $5\times$ and $1/5\times$) is sufficient to capture the stabilization behavior observed across all systems studied.

For the 4-qubit H$_2$ Hamiltonian studied here with $P = 40$ variational parameters, we obtain $\sum_i |c_i| \approx 2.8787$, yielding $\lambda_{\text{scale}} \sim 0.072$.

For the more complex LiH Hamiltonian with $P = 80$ variational parameters, we obtain $\sum_i |c_i| \approx 7.0858$, corresponding to $\lambda_{\text{scale}} \sim 0.089$.

For the RFIM benchmark with $P = 12$ variational parameters, we obtain $\sum_i |c_i| \approx 15.219$, yielding $\lambda_{\text{scale}} \sim 1.27$. In this case, the strongly nonconvex landscape and the shallow single-layer ansatz lead to effective stabilization at smaller $\lambda$ values than suggested by $\lambda_{\text{scale}}$, illustrating that the heuristic should be interpreted as a conservative reference and evaluated in conjunction with system-specific landscape properties.

While these values do not represent optimizer-specific optima, they serve as energy-consistent, order-of-magnitude reference scales that indicate the regime in which regularization effects become relevant. We emphasize that $\lambda_{\text{scale}}$ is not intended to predict the optimal regularization strength, but rather to provide a conservative reference point that should be evaluated together with values moderately above and below it.

Consistent with this interpretation, the effective regularization strength also depends on the target accuracy threshold: tighter accuracy requirements shift the optimal $\lambda$ toward larger values, reflecting the need for stronger landscape conditioning, while looser thresholds permit smaller $\lambda$ values that preserve greater exploratory freedom.

In addition, the regularization coefficient evolves according to a cosine-decay schedule during Stage~A:
\[
\lambda(t) = \frac{\lambda_0}{2}\left[1 + \cos\!\left(\frac{\pi t}{T_A}\right)\right].
\]
The decreasing $\lambda$ schedule is used purely as a heuristic protocol and is not interpreted as a physically or algorithmically optimal strategy.

Across all systems studied, the optimal regularization strength depends systematically on the target accuracy threshold. Looser chemical-accuracy criteria favor smaller values of $\lambda$, permitting greater exploratory freedom and yielding higher success rates at coarse tolerances. In contrast, increasingly stringent thresholds favor larger $\lambda$, reflecting the need for improved conditioning and stabilization of the variational landscape to support fine-scale descent toward the minimum. Operationally, larger $\lambda$ values increase the probability of convergence at tight thresholds (e.g., $10^{-6}$--$10^{-7}$~Ha), while smaller $\lambda$ values achieve higher success rates at looser thresholds but fail to reliably reach high precision. This behavior reveals an intrinsic tradeoff between stabilization and expressivity, indicating that $\lambda$ should be selected based on the target accuracy rather than treated as a universal constant.

\section{Conclusion}

In this work, we have systematically investigated the role of classical regularization in Variational Quantum Eigensolver (VQE) optimization across chemically and physically distinct Hamiltonians, including H$_2$, LiH, and the random-field Ising model. Across all systems studied, the introduction of an $L_2^2$ penalty term produces a consistent, nonmonotonic dependence of optimization performance on the regularization strength $\lambda$, revealing a well-defined stabilization window.

At small values of $\lambda$, the optimization remains largely unconstrained and exhibits high variance and sensitivity to initialization. At intermediate values, regularization stabilizes the variational landscape, leading to higher success probabilities and tighter energy distributions. Excessively large values of $\lambda$, however, over-constrain the optimization and limit the attainable accuracy. Importantly, the location of the optimal regularization strength is not universal, but depends systematically on the target accuracy threshold: looser chemical-accuracy criteria favor smaller $\lambda$, while increasingly stringent thresholds favor larger $\lambda$, reflecting a tradeoff between exploratory freedom and landscape conditioning.

Notably, these qualitative trends persist across different classical optimization algorithms: while conjugate-gradient methods were used for the H$_2$ and RFIM instances and L-BFGS for LiH, all systems exhibit the same characteristic stabilization window and nonmonotonic dependence on $\lambda$. This observation suggests that the benefits of classical regularization are largely independent of the specific optimizer employed and instead reflect generic features of the variational landscape.

\section*{Acknowledgment}

Martin Plesch and Ijaz Ahamed Mohammad were supported by the Research and Innovation Authority project No. 09I03-03-V04-00685. Yury Chernyak was supported by the VEGA project No. 2/0055/23. Computational resources were provided by the DEVANA supercomputer under project No. 311070AKF2.

\bibliographystyle{unsrt}
\bibliography{references}

% Author Biographies
\begin{IEEEbiography}[{\includegraphics[width=1in,height=1.25in,clip,keepaspectratio]{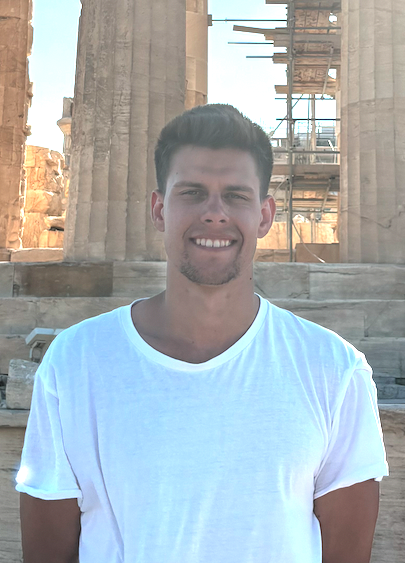}}]{Yury Chernyak} received a B.A. degree in Mathematics and Physics at Hartwick College, and his M.S. degree in Physics from SUNY University at Albany in 2022. He began his doctoral studies at the Institute of Physics, Slovak Academy of Sciences in 2023.
\end{IEEEbiography}
\vspace*{-6em}
\begin{IEEEbiography}[{\includegraphics[width=1in,height=1.25in,clip,keepaspectratio]{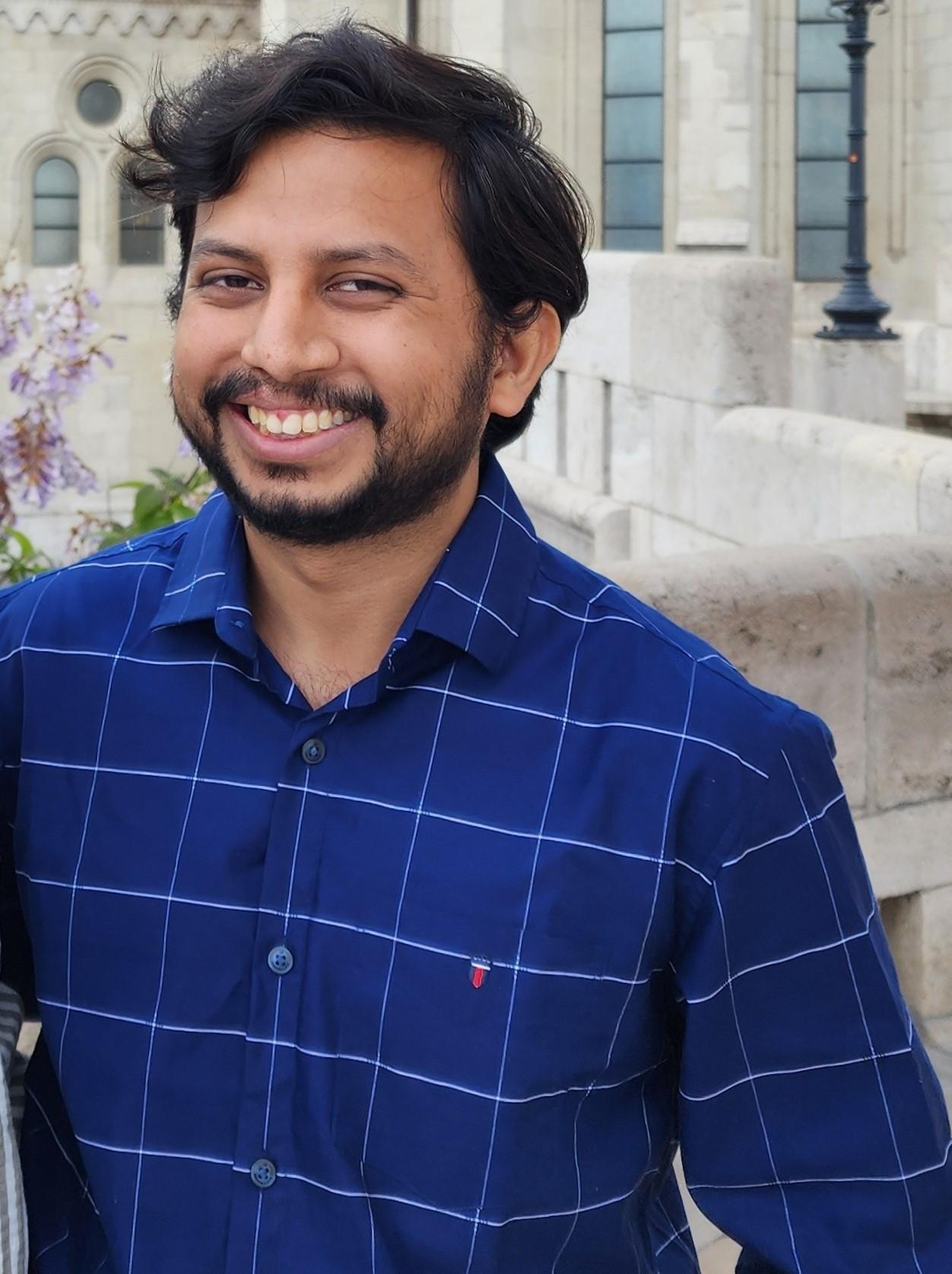}}]{Ijaz Ahmed Mohammad} 
is a senior Physics PhD student specializing in quantum computation. He finished his B.S and M.S in Physics at Indian Institute of Science Educational Research (IISER), Mohali, India in 2021. He was an INSPIRE-SHE scholarship recipient from 2016-2018. He started his doctoral studies at Institute of Physics, Slovak Academy of Sciences in 2021. He has contributed two research articles in the field of quantum information and computation.
\end{IEEEbiography}
\vspace*{-6em}
\begin{IEEEbiography}[{\includegraphics[width=1in,height=1.25in,clip,keepaspectratio]{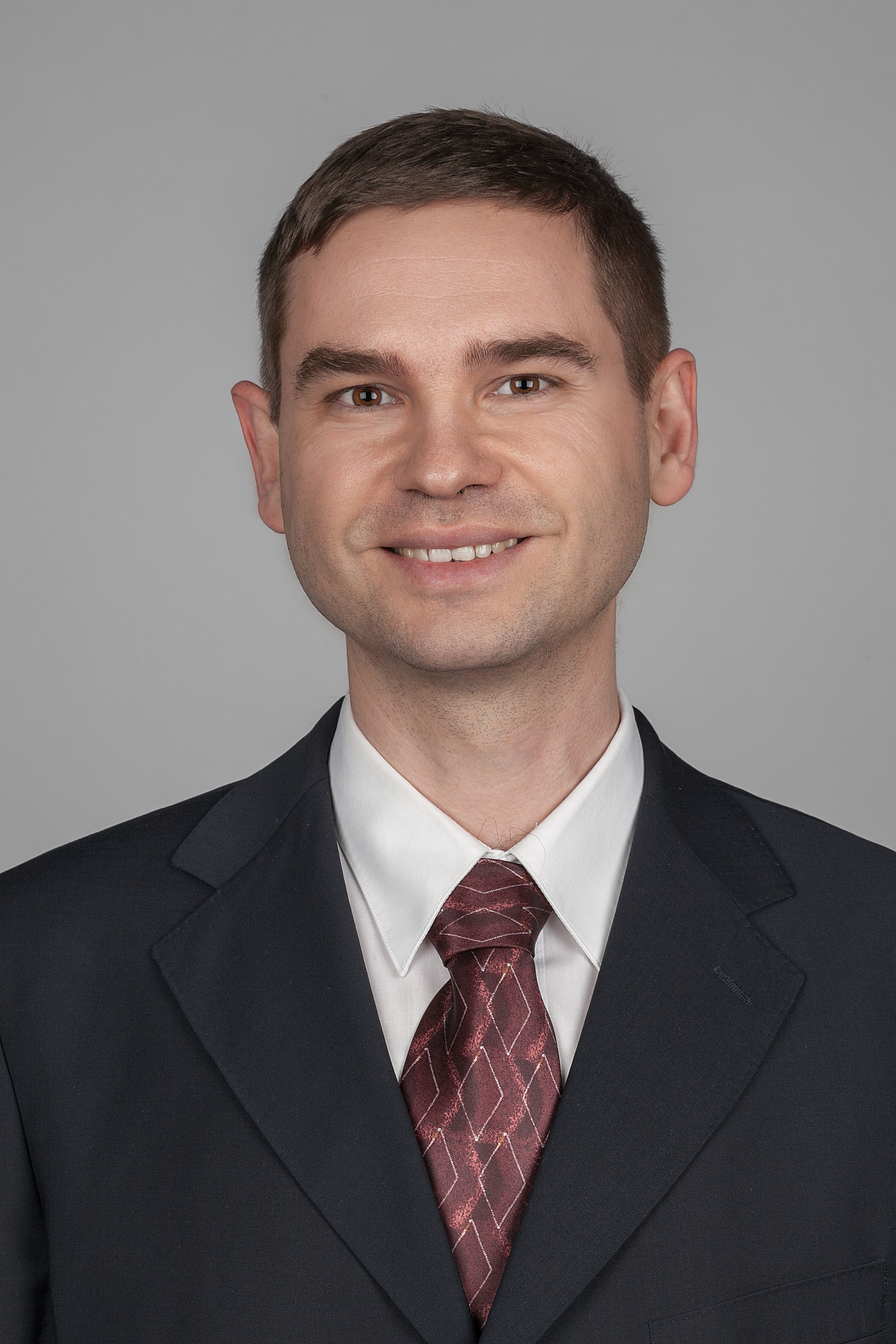}}]{Martin Plesch} is a physicist specializing in complex physical systems and quantum information theory. He is an independent researcher at the Institute of Physics Slovak Academy of Sciences and a Professor at Matej Bel University in Bratislava. With a PhD from the Institute of Physics Slovak Academy of Sciences, Prof. Plesch has held significant roles, including Head of the Department of Complex Physical Systems and Marie Curie Fellow at Masaryk University Brno. He has received numerous accolades for his research and educational contributions, including the Prize for Popularization of Science and the ``Social Innovator'' award. Prof. Plesch is also actively involved in international scientific committees and educational initiatives, serving as a President of the International Young Physicists' Tournament and the World Federation of Physics.
\end{IEEEbiography}

\end{document}